\magnification=\magstep1  
\vsize=8.5truein
\hsize=6.3truein
\baselineskip=18truept
\parskip=4truept
%\parindent=0.3truept
\vskip 18pt
\def\today{\ifcase\month\or January\or February\or
March\or April\or May\or June\or July\or
August\or September\or October\or November\or
December\fi
\space\number\day, \number\year}
\centerline{\bf Balance of forces in simulated bilayers.   }
\medskip
\centerline{by }
\centerline{J. Stecki }
\vskip 20pt
\centerline{ Department III, Institute of Physical Chemistry,}
\centerline{ Polish Academy of Sciences, }
\centerline{ ul. Kasprzaka 44/52, 01-224 Warszawa, Poland}
\bigskip
\centerline{\today} 
\medskip
\centerline{ Dedicated to Professor Keith Gubbins on the occasion of his 70th birthday} 
\vskip 60pt

\centerline{\bf{Abstract}}
Two kinds of simulated bilayers are described and the results are
reported for lateral tension and  for partial 
contributions of intermolecular
forces to it. Data for a widest possible range of areas per surfactant head,
from tunnel formation through tensionless state, transition to floppy 
bilayer, and its disintegration, are reported and discussed. The  
significance  of the tensionless state, is discussed.  


\vfill\eject

{\bf I. Introduction: two-dimensional sheets in three dimensions.}

Great interest in the properties of membranes and bilayers is reflected
in the very large volume of work, reporting simulations of a 
variety of models and simulation methods. However, simulations 
including a range of areas are relativery rare and those including a 
widest possible range of areas - rarer still$^{1-6}$. Most papers concentrate on the 
"tensionless state" for which the lateral tension vanishes. 
 
 The latter is the direct counterpart of the surface (interfacial) 
 tension between two liquids and in fact it is computed according to 
 the same Kirkwood-Buff formula$^{7}$.

At this stage we must mention that the physico-chemical properties of 
a sheet of surfactant molecules, forming a bilayer, are
positively exotic.  Compare with a planar inerface between e.g. 
a liquid and its vapor or two immiscible liquids. Then
the surface tension $\gamma$ is defined by 
$$  (\partial F /\partial A )_V = \gamma  \eqno(1.1) $$
where $F$ is the free energy, $A$ the area, $V$ the volume and the 
constancy of particle numbers and temperature $T$ is understood. But 
$\gamma$ itself is  independent of the area; it is a material property,
a function of state. In a shocking contrast to that,
the  lateral tension of a bilayer (again defined after (1.1))
is area-dependent; moreover in the same system as to composition, density,
and temperature, $\gamma(A)$ can be positive  or negative. Its zero 
defines the "tensionless state" which is of particular interest and
some physical significance.
 
Other properties, such as internal energy $U$ or correlation functions
including the structure factor $S(q)$, also display the area dependence,
if a sufficiently wide interval of areas is investigated. 

The peculiar shape of the {\it bilayer isotherm} i.e. of the function and
plot of  $\gamma(A)$, as shown and discussed in Section II, raises the 
question as to how it originates. A partial answer to that is provided by 
the split of $\gamma$ into its components.

Not all simulations are "atomistic" i.e. not all simulations construct
the intermolecular total energy $E(\{{\bf r}\})$ from 
model intermolecular potentials depending only on the 
spatial coordinates of the constituent particles (atoms or molecules), which
energy is then used in a Monte-Carlo or Molecular Dynamics simulation 
scheme.  One of the advantages of an atomistic simulation is the possibillity
of examining the role of the constituent components such as surfactant 
heads, solvent molecules, and surfactant tails. This is put to use in 
this paper, in which we  report the split of lateral tension.

Our results are reported for two kinds of simulated bilayers; these are
  defined  in Section II. 

In Section III we show the split of lateral tension into components
and we discuss the physical meaning of the "tensionless state".

 In Section IV we return to the discussion of bilayer properties 
and deeper distinctions between interfaces and  membranes or bilayers.
The discussion introduces the concept of constraint.

 It also appears to be necessary to point out that a very important 
category of objects, called "vesicles", is entirely outside the realm
of simulated bilayers. Merging "membranes and vesicles" in one sentence
greatly oversimplifies matters, because vesicles, like spherical micelles
floating in a solvent, enclose a finite volume.  The bilayers do not.

The "widest possible range of areas" of simulated  bilayer, mentioned 
above, refers to the limits of existence of a stable bilayer. When the 
imposed area is 
extended too much, the bilayer recedes creating a spherical hole or rather 
a tunnel filled with solvent  joining the two sides of the solvent 
volume. When compressed, the bilayer flips into a "floppy" state 
(see Section III) which then if compressed further, disintegrates into 
structureless object(s) which cannot be called a bilayer any more.

\bigskip
{\bf Section II. Two kinds of bilayers.}

  A planar bilayer is formed by amphiphilic surfactant molecules made each
 of a hydrophilic polar "head" and hydrophobic non-polar hydrocarbon 
 chains as "tails". The tails form the center of the bilayer and the
 two outer layers of heads separate the tails from 
 the polar solvent (water). Therefore the cohesive energy density (CED)$^{8}$ 
 is high (solvent), high (first layer of heads), low (tails), low  
 (second layer of tails), high (second layer of heads), high (solvent on
 the other side of the bilayer). 

The same forces and the same preferences 
 operate in the formation of micelles of various shapes. 
 Normally the solvent is polar, of high CED, most commonly water, 
and the micelles shield their hydrocarbon tails by an outer layer of 
heads in contact with water. However, there exist rare examples of 
{\it reverse} micelles which are formed by amphiphilic molecules in a
non-polar solvent of low CED. In reverse micelles, the (hydrocarbon) 
tails form the outward shell and the (polar) heads are in the center 
of the micelle. By analogy with these I have$^{2,4}$ constructed in simulations
{\it reverse bilayers} which are formed in a non-polar solvent of low 
CED. On crossing the reverse bilayer along the z-direction perpendicular to
the x-y plane of the bilayer, the sequence is now: solvent, tails, heads,
heads, tails, solvent or, in terms of CED : low,low,high,high,low,low.

   These cases can be and have been modelled with the use of Lennard-Jones
(6-12) potential (LJ) with the minimum of 3 kinds of particles: s(solvent),
h (head), t (tail). With 6 energies, 6 collision diameters, and 6
cut-off radii, various 
 simplifications have been used  in bilayer simulations   with the LJ 
potentials$^{1-4, 5, 9-13}$.

 In our work 
 the solvent was made of structureless spherical particles; freely jointed
 chains of such particles of the same size have modelled the
 surfactant molecules making up the bilayer. 

 We found certain regularities in the stability of the model bilayers.
 We found that the chain lengths of the tails can be shortened even down to 
 unity, the molecule becomes then a dimer ("h-t") made of two particles.
 We also found that it is worthwhile to keep the 
 presence of the solvent;  in some simulations
 very unusual, in fact unphysical, intermolecular forces  
 were required in order to ensure existence (in simulation) of stable 
 bilayers$^{14}$ made of trimers. The {\it reverse} bilayers 
 made of trimers were successfully 
 simulated without a solvent $^{15}$, although it appears this was not 
 followed with further work.
 We also found that longer chains stabilize the bilayer; of the tail 
 lengths ${l} =1,4,8$ the dimers were difficult to stabilize without
 an extra repulsion (replacing the hydrophobic effect) from the solvent 
whereas, for longer chains, this extra
 repulsion was not necessary for stability. Chains with $l=8$ produced 
 stable bilayers with great ease. 

The modelling of solvent as made of spherical structureless particles 
interacting with a central potential creates a certain conceptual difficulty 
because of the hydrophobic effect. 
It has been partially resolved by an introduction$^{9}$
of an additional repulsive force between the solvent and the tails, e.g.
of the form $c/r^n$ with $c>0$ and n=10 or more$^{9-13}$. A 
temperature dependence would be needed too, to take into account 
the entropic effects of the hydrophobicity. 

In the case of a reverse bilayer, the reasons for its formation are 
mostly energetic: a pair of heads (now inside the bilayer) attract 
each other more strongly than any other pair 
and there is no hydrophobic effect since the 
solvent is non-polar with a weak CED.

\vfill\eject

{\bf Section III. The Lateral Tension and its Components.}

A bilayer originally planar, lies in the $x-y$ plane in the middle 
of a simulation box which is a parallelepiped of volume $V=L_x\times 
L_y \times L_z$; its area is $A=L_xL_y$. It is made of $N_m$ molecules
which contain $N_m$ heads, originally $N_m/2$ in each of the two 
monolayers, so that the "area per head" $a=A/(N_m/2)$.   The 
bilayer is surrounded "from above" and "from below" by $N_s$ 
molecules of the solvent. The overall density is $\rho = N/V$ with 
$N = N_s+(l+1)N_m $. The lateral tension is defined$^{7}$ as the response 
to the process of increasing the area at constant volume. With 
$L_x \rightarrow L_x+\delta L_x $, $L_y \rightarrow L_y+\delta L_y $,
$L_z \rightarrow L_z+\delta L_z $, the constraint of constant volume 
requires $ \delta L_x/L_x + \delta L_y/L_y + \delta L_z/L_z =0$
and the Kirkwood-Buff equation follows
$$ \gamma A =L_z (2p_{zz}-p_{xx}-p_{yy})/2 \eqno(3.1)$$
with the known definitions 
$$  p_{\alpha\alpha} = 
   <\sum_j r^\alpha_j\dot {{\partial E}\over{\partial r^\alpha_j}} > \eqno(3.2)$$
valid for a rectangular box. The average is a canonical 
average at given temperature $T$, all $N'$s, $A$, and $V$.
In the simulation, for each state point of $T,V,N_m,N_s,A$ we obtain
one value of $\gamma (A)$. In our work, $L_x=L_y$; changing $L_x$ to 
any new value, we calculate the new value of $L_z$ needed to keep the volume
$V=L_xL_xL_z$ at exactly the same value. 

Each time $\gamma$ is computed, it is  a sum of binary terms and these 
are grouped into partial sums so as to produce the split 
$$ \gamma = \gamma_{ss} + \gamma_{sb} + \gamma_{bs} + \gamma_{bb}. \eqno(3.3)$$
Here the indices $ss$ refer to the solvent-solvent part, the indices 
$sb$ and $bs$ to the solvent-bilayer part, and $bb$ refers to the bilayer-bilayer part. There is no approximation involved.

Fig.1 shows an example of a bilayer made of chain molecules of 
tail length $l=4$
plus one more particle as the head, immersed in a solvent at high liquid-like
density. Each $\gamma_{\alpha\beta}$ is plotted against the area per head
$a$; the lines joining the data points are there to guide the eye, only.
The contribution of solvent-solvent pairs is about 0.2-0.3  and almost constant
i.e. independent of $a$. All pairs made of solvent particle and any particle
of any bilayer molecule make up the $sb$ or $bs$ contribution; it is also
positive, varying smoothly with $a$ and taking values between 0.5 and
1.05 .
The $bb$ contribution shows the characteristic break into two curves, 
nearly linear with $a$, but one with positive slope and the other with a
negative one. This pattern is transferred to the sum,  eq.(3.3),
total $\gamma(a)$.

Fig.2 shows the effect of size for the same chain-length $l=4$. 
The solvent-solvent contribution $\gamma_{ss}(a)$ is practically independent 
 of size, as is nearly so the solvent-bilayer part. The $bb$ part and,
 consequently, the total $\gamma$ display the break into two regions;
the floppy part  strongly affected by size and the extended region
negligibly so. The bigger size shows flatter $\gamma(a)$, still negative,
but closer to zero. Lowest values of $\gamma(a)$, which may be taken as
the border between the floppy region and the extended region, are negative
but lower value corresponds to smaller size.
 
 These remarkable patterns are repeated for chain length $l=8$, with
breaks much more pronounced. Fig.3 shows $\gamma_{\alpha \beta}(a)$ 
for two sizes; the intermediate size is omitted, just as was in Fig.2,
in order to make the plot clearer. Again the $ss$ part is constant and
size-independent, the $sb$ part nearly so, and the total $\gamma$ 
along with the $sb$ part, show these  remarkable features: breaks and 
abrupt changes of slope. For 
large $a$ the slope is positive and the size-dependence negligible.
The transition to the floppy region occurs near $a=1.8$ for the smaller
system and the slope $d\gamma /da$ changes sign; the break is very 
pronounced. The  transition is shifted to $a=1.9$ in the bigger system,
the break is more pronounced in the $bb$ part than in the total $\gamma$.
Most characteristically in the bigger system $\gamma(a)$ is flat and 
near zero (though still always negative) in the floppy region.
The tensionless point appears to lie on the r.h.s. curve ( extended bilayer
region) in this case. 
 
All bilayer isotherms i.e. all plots of $\gamma(a)$ we have obtained$^{1-4}$
show the same pattern of two curves joined at some negative value of 
the lateral tension $\gamma$.  We interpret the 
region of lower $a$ as the region of the {\it floppy state} of the 
bilayer which escapes into the third dimension,  buckles, and takes a
fuzzy and irregular form.
 A gently fluctuating planar bilayer is found in the 
other region at higher $a$'s. Generally, the region
of extended bilayer includes all states with positive 
$\gamma$. The crossing of the isotherm with the $a$-axis marks the
{\it tensionless state} . In the plots of Fig.1-3 the  region of the 
extended bilayer has a large positive slope of $\gamma(a)$.
Such positive slopes have been found independently$^{5,6}$
 and also earlier$^{9-13}$. The entire range including the floppy region
is found only in few references$^{2-6}$. The corresponding split of the
lateral tension is found in this paper only.

The status of the tensionless state appears now in a new light, as an 
accidental event resulting from the algebraic sum of positive and 
negative contributions after (3.3). The partial contributions suggest 
that the break point dividing the two branches of $\gamma(a)$ (or 
$\gamma_{bb}(a)$ ) is a truly characteristic point which ought to have 
a physical significance, rather than $\gamma(a_0)=0$.

  The shortest possible tail of one particle is present in a dimer. We 
have attempted and partially succeeded$^{1}$ in simulating bilayers made 
of dimers, but these appeared to be only imperfectly stabilized
by strong head-head and head-solvent interactions or/and the
extra-repulsion introduced artificially to mimick the hydrophobic effect,
as invented earlier$^{9}$. Possibly at other densities and temperatures
the simulations would have been more encouraging. We do not show these
results. As mentioned in Section II, we turned to the new case of {\it reverse}
bilayers formed in a weakly interacting solvent and succeeded in producing
(in simulation) stable reverse bilayers of amphiphilic dimers$^{2,4}$.
Fig.4 and 5 show a selection of several characteristic cases, each for two sizes.
These cases are: (a) extra repulsion added, (b) no extra repulsion,
(c) no extra repulsion and longer range of attractive forces. In cases
(a) and (b) the cut-off was the generally used $r=2.5$, in the case (c)
it was $r=3.2$. For clarity, we split the data into two Figures: Fig.4
shows $\gamma$ total only, for all 5 systems whereas Fig.5 shows the
components $ss$, $sb$, and $bb$. All curves for all 5 systems in Fig.4
show a pattern similar to $\gamma(a)$ in Fig.3  (for $l=8$). This 
characteristic pattern is : almost linear rise with $a$ for larger 
$a$ and almost constant negative $\gamma$ for small $a$. 
Such a description fits bigger systems better. The biggest system of 
an area about $100\times 100$ of reverse dimers case (a) is shown in 
Reference 4.
As can be seen
from Fig.4, the plot  again strongly suggests a transition between two forms
of the bilayer and certainly the existence of two regions$^{1-5}$. 
 In the floppy region 
$\gamma < 0$ and the bigger the system  the closer to zero $\gamma$ becomes.
Fig.5 shows the components of $\gamma$ but without $\gamma$ itself.
The pattern is similar to those shown in Fig.1,2,3 except that the 
$ss$ contribution to $\gamma$ is now negative, between -0.6 and -0.2,
the $sb$ contribution much larger, between 2. and 3.6. The $bb$ 
contribution is, for all sizes, practically linear in $a$ with again
negative slope and negative value in the region of "floppy" state. 
  The transition appears softer than for chain molecules, especially
those with longer tails where the breaks were sharp (see Fig.3) in 
small systems.

   The status of the tensionless state appears to be relegated to
the category of accidental coincidences. An analogy may be drawn with 
the equation of state of liquid in equilibrium with its saturated 
vapor; certainly $P=1$ is a particular point on the vapor pressure 
curve, but it has no particular physical meaning - in contrast with e.g. the
triple point. The only physical consequence of zero in lateral tension 
is the special form of the structure factor$^{10,1-4,5}$. However, even 
this has been questioned recently$^{12,4}$. In a detailed investigation
of the structure factor $S(q)$ of a simulated bilayer, which 
should, for small $q$,
asymptotically vary as $1/S \sim k q^4 + g q^2 $ where 
$k$ is the rigidity constant and $g$ - the surface tension, we found
 $g \ge \gamma $ systematically$^{4}$.  The explanation
 advanced elsewhere$^{12}$ is that $\gamma \delta A$
is not the correct work of deforming a bilayer initially planar, but 
$g \delta A_{true}$ is. Here $A_{true}$ is the true area of the interface
or bilayer, obtained by following its surface, whereas $A\equiv L_xL_y$ is
the {\it projected area}. Since $S(q)$ measures the spontaneous fluctuations,
$g q^2 $ is correct and $\gamma q^2$ is not. On the other hand, when 
measuring the response of the system to an imposed  change in the projected area, 
$\gamma \delta A$ is the correct free energy increment. 
The increase in the projected area is under control of the experimentalist;
the change of the true area induced by the twists and bends of the 
bilayer, is not. By taking these arguments into account, the correct definition of 
the tensionless state was assumed$^{12}$ to be $g=0$ and interpolation procedures
were used$^{12}$ to determine the new correct tensionless state.
Whether the details of the advanced explanation$^{12}$ are correct or not,
it is obvious that $g$ parametrizes the spectrum $S(q)$ of spontaneous
shape fluctuations. 
The difference between $g$ and $\gamma$ was also discussed 
elsewhere$^{13,6,15}$.
The zero of $\gamma(a)$ is unambiguous in small systems; in large system
of reverse bilayers$^{4}$ it falls within the narrow region of fastest
change in slope, signalling the transition of the bilayer to the
floppy state. Then the precision of its determination is doubtful. 
Some of this 
ambiguity is seen in Fig.4 and in Figures of Reference 4.

{\bf Section IV. Discussion.}
 
We have noted in the Introduction that, to a physical chemist familiar 
with the interfaces, some properties of a bilayer are truly exotic. 
Let us list these properties: 
~~ (1) the derivative given by eq.(1.1) depends on the 
specific area; 
~~ (2) it can be negative; 
~~ (3) extension of the area leads to a
hole or tunnel as the bilayer resists further extension; 
~~ (4) there is 
distinction between actual area and the projected area but the bilayer
keeps its true area nearly constant; 
~~ (5) as there are two areas, there are two interfacial tensions.
 
Definitely this is not what we learn from classical textbooks on 
thermodynamics.

The feature (1) has been seen in all simulations, including these
over a wide range of (projected) areas$^{1-6,9-15}$. The feature (2)
has also been seen$^{2-6,11,12,13,15}$. The feature (3)
has not only been seen$^{2,4,15}$ but also investigated$^{15}$. Distinction
 (4) has been the subject of papers$^{13,5}$ specifically devoted to it. 
Finally, the existence of an(other) surface tension(s), besides the lateral 
tension, has been repeatedly surmised$^{5,6,11,13}$; equations have even 
been derived$^{14}$.

The explanation of such specific behaviour
ultimately lies with intermolecular forces and steric effects which any 
future theory will have to take into account.
The bilayers are made of {\it amphiphilic} molecules, each containing
a hydrophilic part and a hydrophobic part (or more generally, "solvophilic" and
"solvophobic"). These molecules use their very special setup of intermolecular
forces to create a stable sheet of constant mass (and approximately constant
area). This creates a new situation with a constraint of given total area 
almost conserved when the projected area is varied 
 by varying the simulation box.

The constraint of constant mass (or anything related to it) is entirely 
absent in fluctuating interfaces.
The molecular mechanism of fluctuations of shape of e.g. a liquid-vapor 
interface, involves diffusion from either phase; a local protrusion or 
excursion of 
the interface takes place not only by deforming the existing 
surface but also by absorbing or releasing 
molecules from/to either phase. 

 The surface of a perfect crystal in equilibrium with its vapor 
is a good example; all shape fluctuations are due to 
evaporation or condensation. 

Such 
processes clearly take part in shape fluctuations of a liquid surface as
well.

Hence an interface is a system open with respect to particle number. 
It is not so with the bilayer. Not only the solvent is 
virtually absent from 
the bilayer, but also the surfactant molecules making the bilayer never leave
it. In principle the surfactant must be present in the solvent
but  the bilayer changes shape by changing
the position of its molecules, not by exchanging them with  
 the solvent solution. The shape fluctuations take place {\it under the 
constraint} of constant number of particles, i.e. constant mass.

A constant mass is implicit in the elastic theory
of solids and the links to that field have already been explored$^{5,10}$.
Conversely, application of the elastic theory of solids to liquid
interfaces, (the so called "drumhead model"), is clearly not quite correct
by not allowing for diffusion described above. 

We discuss now the new evidence for the transition.
All plots of $\gamma_{bb}(a)$ and $\gamma(a)$ show some kind of break 
into two branches, one corresponding to the "floppy" bilayer. 
Although the word "floppy" or "buckling" appeared occasionally$^{4,5}$,
 in fact 
the status of the floppy region and of the transition into that region
are not very clearly established. 
There are suggestive analogies with the crumpling 
transition of solid (tethered) membranes or with coiling transitions of 
polymers. This would suggest it is a first-order transition.
The breaks in $\gamma_{bb}(a)$ (see Section III) strongly support such 
a hypothesis also suggesting it is an internall reorganization of the 
bilayer structure, in which the solvent plays a secondary role. 

Further support is found with the radius of gyration of the bilayer,
shown in Fig.6; the break in slope is very clear. 

Finally, the issue of the other surface tension is not fully resolved.
As we discussed above in Section III, one may define two surface tensions,
one coupled to the projected area and the other coupled to 
something else, perhaps the true 
area. The first one is the lateral tension and the other 
may be that given by the virial expression derived or rather proposed
recently$^{14}$ or $g$  parametrizing the structure factor $S(q)$. 
(See the end of Section III). These last two may be equal or may be not.

 I hypothesize that $g$ may be related to that given 
by the Zwanzig-Triezenberg (Z-T) equation$^{16,7}$ - because 
the latter is derived by considering a spontaneous fluctuation.
Now, in simulations of liquid-vapor interface, 
I have determined$^{17}$ the surface tension from 
both expressions and these two numbers were equal within 0.01 percent. 
Thus I have confirmed the validity of the Schofield proof$^{7}$ of the 
equivalence.
Since the (Z-T) equation is derived in a grand-canonical ensemble, $g$
can only be "related to it" as stated above; the constraint would
have to be introduced.

We finally remark that the constraint appears as a most natural thing 
in description of polymers when the polymer string has a constant number
of segments. Indeed a model  one-dimensional 
membrane/bilayer embedded in two dimensions, 
 is identical to a model of a polymer string on a plane.
\vfill\eject

{\bf Appendix.  Details of the Model Used }.
 
The simulations themselves were in the past and also recently 
 based on the Verlet leap-frog algorithm with Nose-Hoover thermostat and
 were never shorter than half a million steps, about 0.9 million in 
 almost all runs. The intermolecular potential energy was a sum over pairs of
 particles; all pairs were interacting Lennard-Jones 6-12 potentials
 $$ u(r)=4\epsilon\big(~ (\sigma/r)^{12} - (\sigma/r)^6~\big) $$
with different parameters $\epsilon$ and $\sigma$ and with different 
 cutoffs $r_c$. A constant $u_0$ may be added at will and this makes
 no difference to forces. $u(r)$ has a minimum at $r=r^\ast$ of depth 
 $u(r^\ast)=-\epsilon$.
There are two kinds of spherical particles "a" and "b".
 The solvent was made of particles "a" and the amphiphilic molecules
 of chains "abb..b". The collision diameter $\sigma$ was common to all
 pairs. The cutoff parameter was $r_c=2.5\sigma$ for all like pairs
 and $r_c = r^\ast$ for unlike-pair intermolecular force. All potentials
 were cut-and-shifted to assure continuity of force. Molecular Dynamics
 algorithms use forces. 
 
 The bilayer-forming amphiphilic molecules were freely-jointed chains
 and each intramolecular bond was made of two LJ halves thus:
 $$ u_{bond} = u(r) ~~~~  r<r^\ast  $$
 $$ u_{bond} = u(2r^\ast-r) ~~~~ (r^\ast<r<2 r^\ast)  $$
 $$u_{bond}= +\infty ~~~~ r > 2 r^\ast , $$
for either "a-b" or "b-b" bond.
As explained in the text, for the special case of "reverse bilayers"
made of dimers, $a-b$, the "bb" intermolecular attractions were enhanced 
by making $\epsilon_{bb} >> \epsilon_{aa}$. Thus "b"'s became "heads" 
located in the center of the bilayer.

\vfill\eject

{\bf Acknowledgements}.
This paper is
dedicated to Professor Keith Gubbins on the occasion of his 70th birthday. 
The author gratefully acknowledges most useful interactions with 
Professor Keith Gubbins while visiting him at Cornell University. 

\bigskip

\noindent {\bf  REFERENCES  }.

 \item {$^1$} J. Stecki, Intl. J. Thermophysics {\bf 22},175 (2001).
 \item {$^2$} J. Stecki, J. Chem. Phys. {\bf 120}, 3508 (2004).
 \item {$^{3}$}  J. Stecki, J. Chem. Phys. Comm.{\bf 122}, 111102 (2005).
 \item {$^{4}$}  J. Stecki, J. Chem. Phys. {\bf 125}, 154902 (2006).
 \item {$^{5}$}  W. K. den Otter, J. Chem. Phys.{\bf 123}, 214906 (2005).
 \item {$^6$} Hiroshi Noguchi and Gerhard Gompper, Phys. Rev. E {\bf 73}, 021903 (2006).
 \item {$^{7}$}J.~S. Rowlinson and B. Widom, 
  {\it Molecular Theory of Capillarity} (Clarendon, Oxford, 1982), page 89-90.
 \item {$^{8}$}J.~S. Rowlinson, {\it Cohesion} (Cambridge U.P., 2002).
 \item {$^9$} R. Goetz and R. Lipowsky, J. Chem. Phys. {\bf 108}, 7397 (1998).
 \item {$^10$} G. Gompper, R. Goetz, and R. Lipowsky, Phys. Rev. Lett. {\bf 82}, 221 (1999).
 \item {$^{11}$} A. Imparato, J. C. Shilcock, and R. Lipowsky, Eur. Phys. J. E. {\bf 11}, 
    21 (2003);  see also Reference 13. 
 \item {$^{12}$} A. Imparato, J. C. Shilcock, and R. Lipowsky, Europhys. Lett. 
  {\bf 69}, 650 (2005). 
\item {$^{13}$} A. Imparato, J. Chem. Phys. {\bf 124}, 154714 (2006). 
\item {$^{14}$}  O. Farago, {\it ibid.} {\bf 119}, 596 (2003); for 
 further work on this model see 
    O. Farago and P.Pincus,  {\it ibid.} {\bf 120}, 2934 (2004).
\item {$^{15}$} Ira R. Cooke and Marcus Deserno, cond-mat/0509218 (8 Sept.2006);
                  Europhysics Letters,
 \item {$^{16}$} D. G. Triezenberg and R. Zwanzig, Phys. Rev. Lett. {\bf 28}, 1183 (1972). 
 \item {$^{17}$} see  J. Stecki, J. Chem. Phys. {\bf 114}, 7574 (2001); ibid. 
 {\bf 108}, 3788 (1998).

\vfill\eject

{\bf Captions to Figures}
\bigskip

{\bf Caption to Figure 1}

The general pattern of splits of lateral tension $\gamma$ according 
to (3.3) is shown as plots against the area-per-head  $a$; here
 for an intermediate size of simulated bilayer with $N_m=1800$ molecules 
 with tails $l$=4 segments long, with all collision diameters
 $\sigma$ and all energies $\epsilon$ equal , at $T=1.$, 
$N_s+(l+1)N_m=49000$,
liquid density $\rho=0.89204$. The $ss$ contribution is marked with squares,
the $sb$ contribution - with stars, the $bb$ contribution - with crosses , 
and  the sum - with black circles .
All data are quoted everywhere in LJ units reduced by the energy depth $\epsilon$
and collision diameter $\sigma$. The lines are there to guide the eye.

\bigskip

{\bf Caption to Figure 2}

 The split of
lateral tension $\gamma$ plotted against the area-per-head $a$
 for two sizes (SM) and (B) of simulated bilayer, made of $N_m=450$ (SM)
 and $N_m=3973$ (B) molecules 
 with tails $l$=4 segments long, with all collision diameters
 $\sigma$ and all energies $\epsilon$ equal, at $T=1.$, 
 $N_s+(l+1)N_m=12250$ (SM) and 105134 (B),
liquid-like density $\rho=0.89204$. With the notation of eq.(3.3), 
the contribution 
 ${ss}$ is shown with triangles (SM) and squares (B) near 0.2-0.3;
 ${sb}$ - plus signs (SM) and stars (B) near 0.5-0.7;
 ${bb}$ - open circles (SM) and crosses (B) 
 with negative slope for small $a$ and positive slope for large $a$.  
 The sum after (3.3), $\gamma$,
 is shown with thick lines,
 black squares (SM), and black circles (B). See text
 and Caption to Fig.1.  Compare with  Fig.3.

\bigskip

{\bf Caption to Figure 3}

 The split of
lateral tension $\gamma$ plotted against the area-per-head $a$
 for two sizes (SM) and (B) of simulated bilayer, 
made of $N_m=450$ (SM) and $N_m=3973$ (B) molecules 
 with tails $l$=8 segments long, with all collision diameters
 $\sigma$ and all energies $\epsilon$ equal , at $T=1.$, 
 $N=N_s+(l+1)N_m=14050$ (SM) and 121026 (B), density $\rho=0.89204$. 
 With the notation of eq.(3.3), the contribution 
${ss}$ is shown with triangles
(SM) and squares (big) near 0.2-0.3;
 ${sb}$ - with plus signs (SM) and stars (big) near 0.5-1.1;
 ${bb}$ - with open circles (SM) and crosses (big). Note the  
 negative slope for small $a$ and positive slope for large $a$.  
 The sum after (3.3) is shown with thick lines,
 black squares (SM), and black circles (big). See text.
   Compare with  Fig.2 and note: the unique breaks 
 in  $\gamma_{sb}$, almost constant slope in the region of floppy 
 bilayer, translated into zero slope in total $\gamma$.

\bigskip

{\bf Caption to Figure 4}

Total lateral tension $\gamma$ of reverse bilayers made of dimer molecules
plotted against area-per-head $a$ for 
biggest (B) or smallest (SM) size, and 3 cases: 
(a) extra repulsion added and head-head attraction enhanced (H-H); 
(b) only (H-H); 
(c) the cutoff of attraction is a 
  long-range $r_c=3.2$, not $r_c=2.5$; otherwise (a).
The other parameters are: $T=1.9,\rho=0.89024$,   
 $N_m=2238$ dimers, $N=N_s+2N_m=40000$(SM) and
$N_m=5760$ dimers, $N=N_s+2N_m=160000$ (B).
These had areas near $36\times 36$ and near $55\times 55$. The points are: 
open squares - Ba; plus signs - SMa; black squares - Bb; crosses -SMb;
stars -SMc. The lines are there to guide the eye only. The deviation of
data-point at $a\sim 1.13$ (black square) is due to the tunnel (hole).
see Fig.5 and text.   

\bigskip

{\bf Caption to Figure 5}

The split of lateral tension after (3.3) of systems shown in Fig.4. See
caption to Fig.4 for parameters. Here the points are:
case "Ba" contribution $ss$ - open squares, $sb$ - black triangles,
$bb$  - pentagons;
case "SMa" and $ss$ - plus signs, $sb$ - open circles, $bb$ -black triangles;
case  "Bb" and $ss$ - black squares, $sb$ - open triangles, $bb$ -black pentagons;
case "SMb" and $ss$ - crosses, $sb$ - black circles, $bb$ -diamonds;
case "SMc" and $ss$ - stars, $sb$ - open triangles, $bb$ - black diamonds.
All $ss$ contributions cluster near -0.5, all $sb$ are $>2.$, and all
$bb$ show two regions interpreted as floppy bilayer and extended bilayer.
See text. 

\bigskip

{\bf Caption to Figure 6}

One example of the radius of gyration for the same bilayer whose $\gamma$ 
is plotted in Fig.1, also plotted against area-per-head $a$. It appears
to support, along with other correlations$^{3,4}$, the hypothesis of a 
discontinuous transition between the extended and floppy bilayer. 

\vfill\eject

\vfill\eject\end
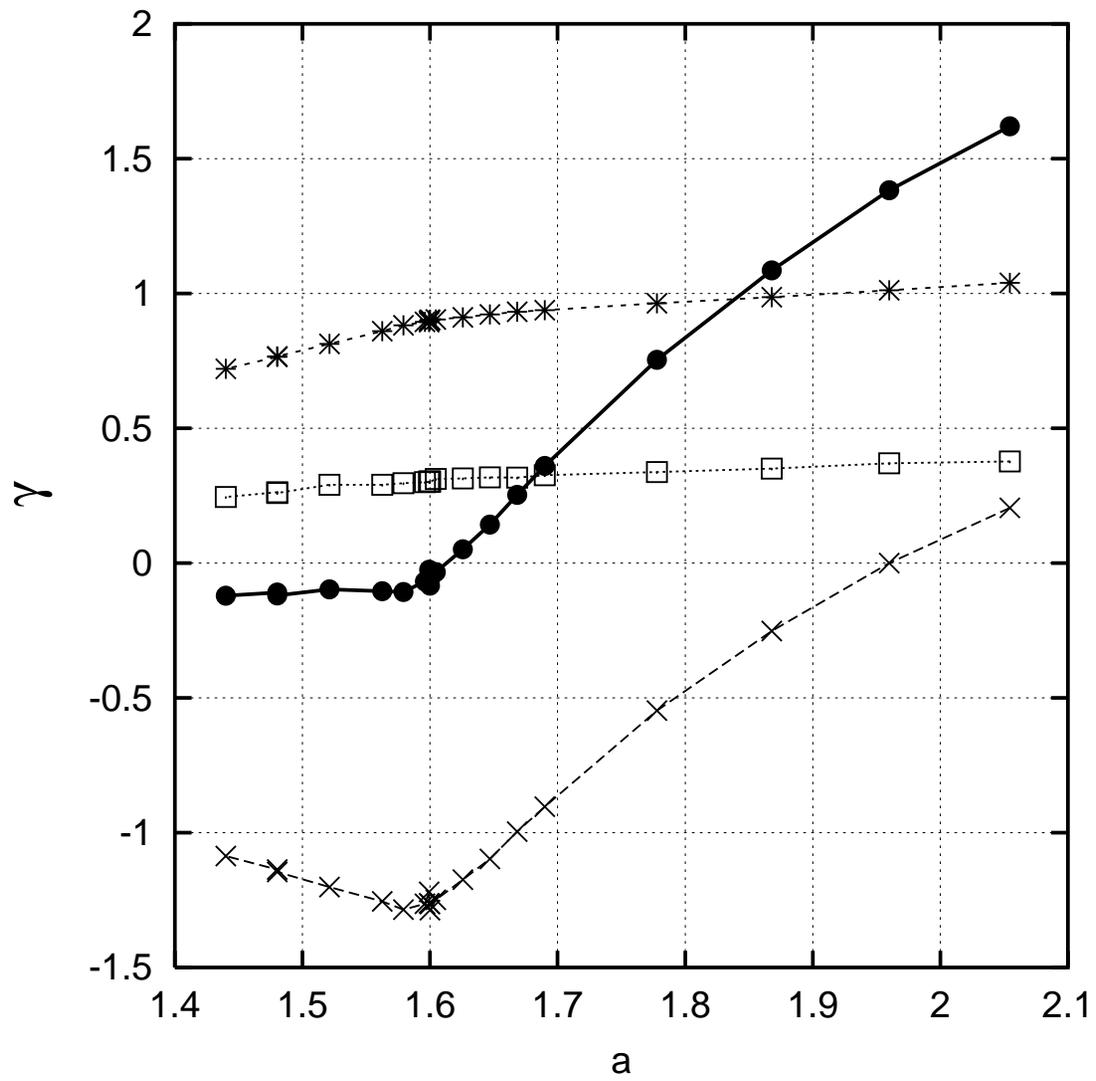
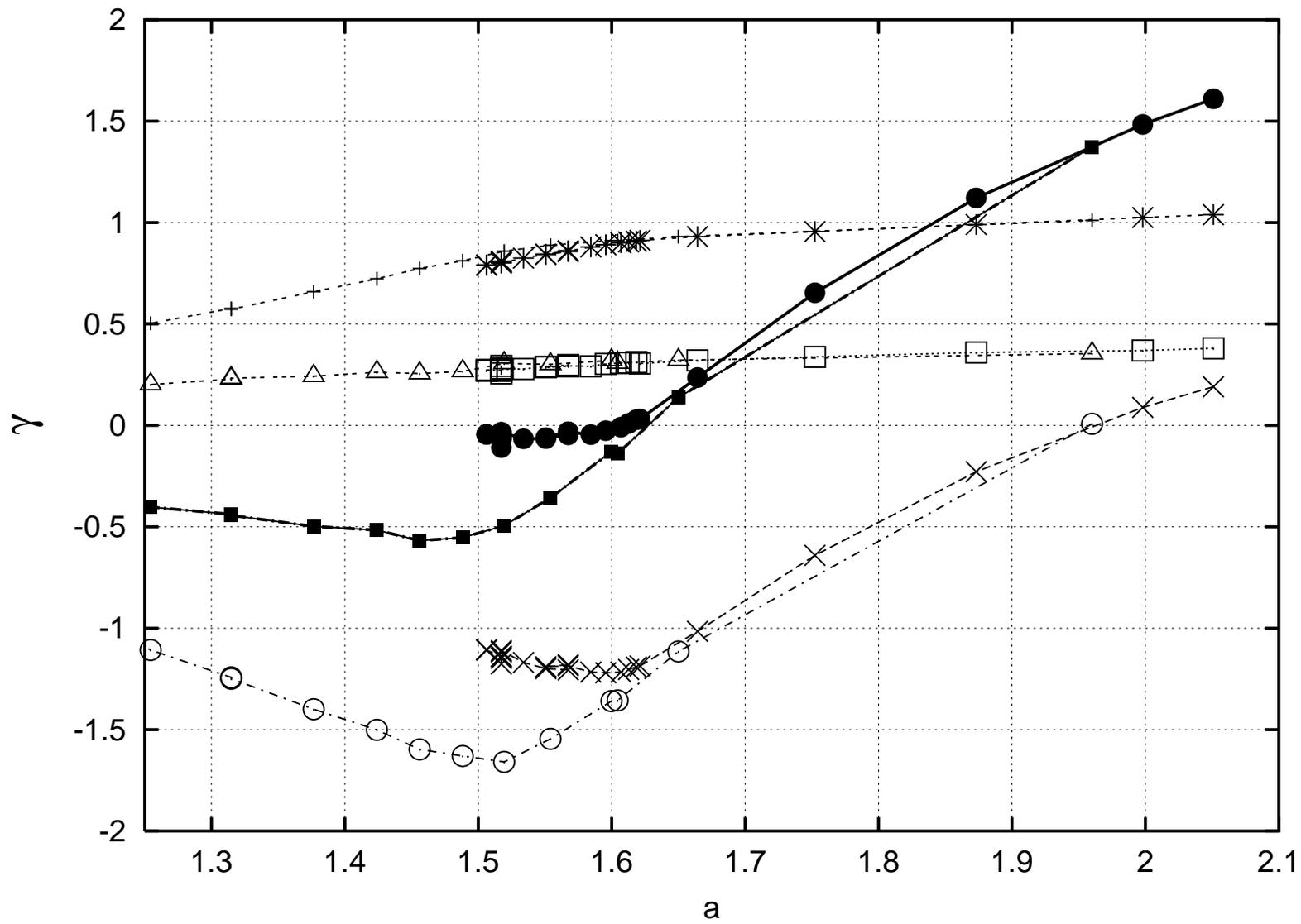
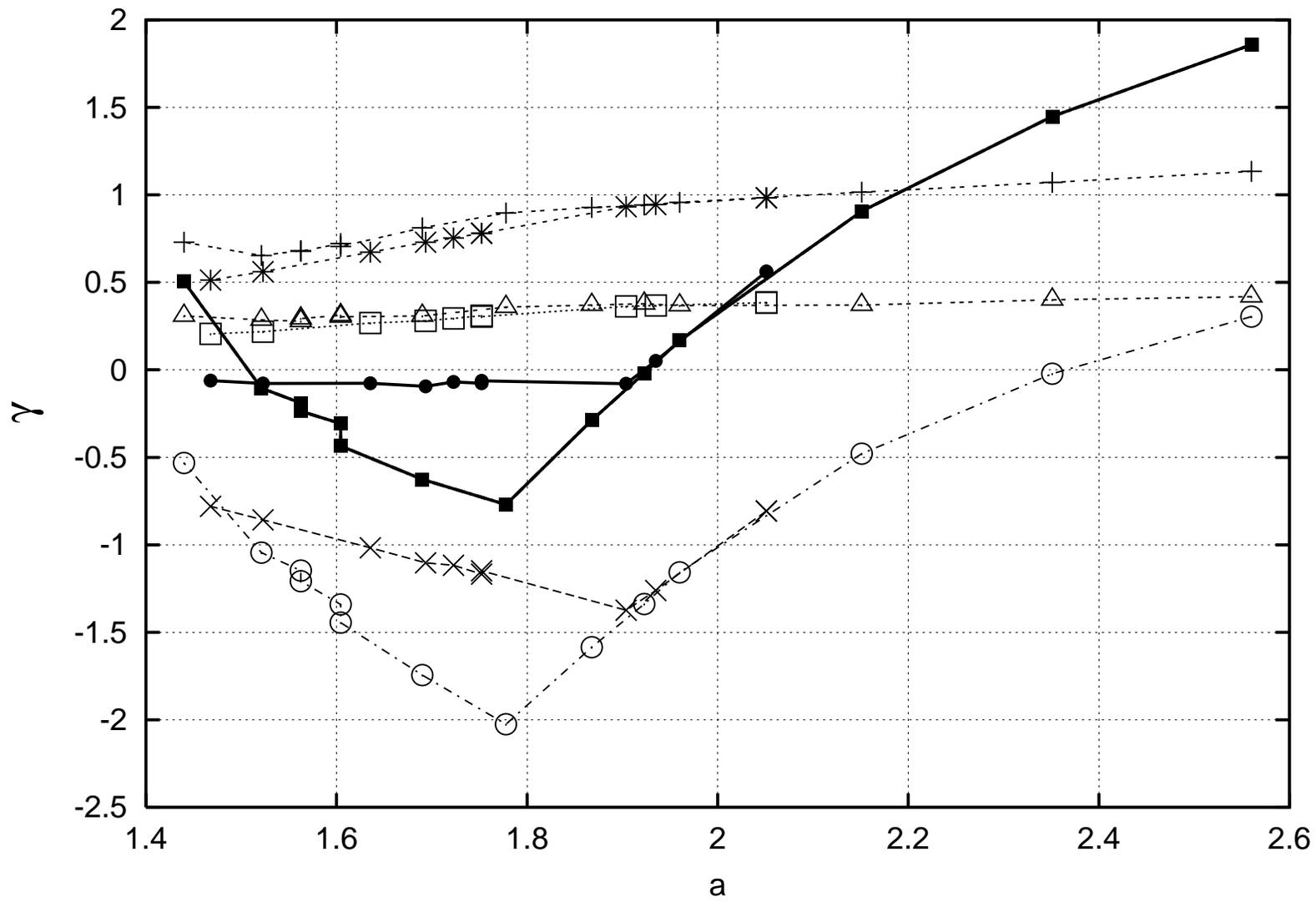
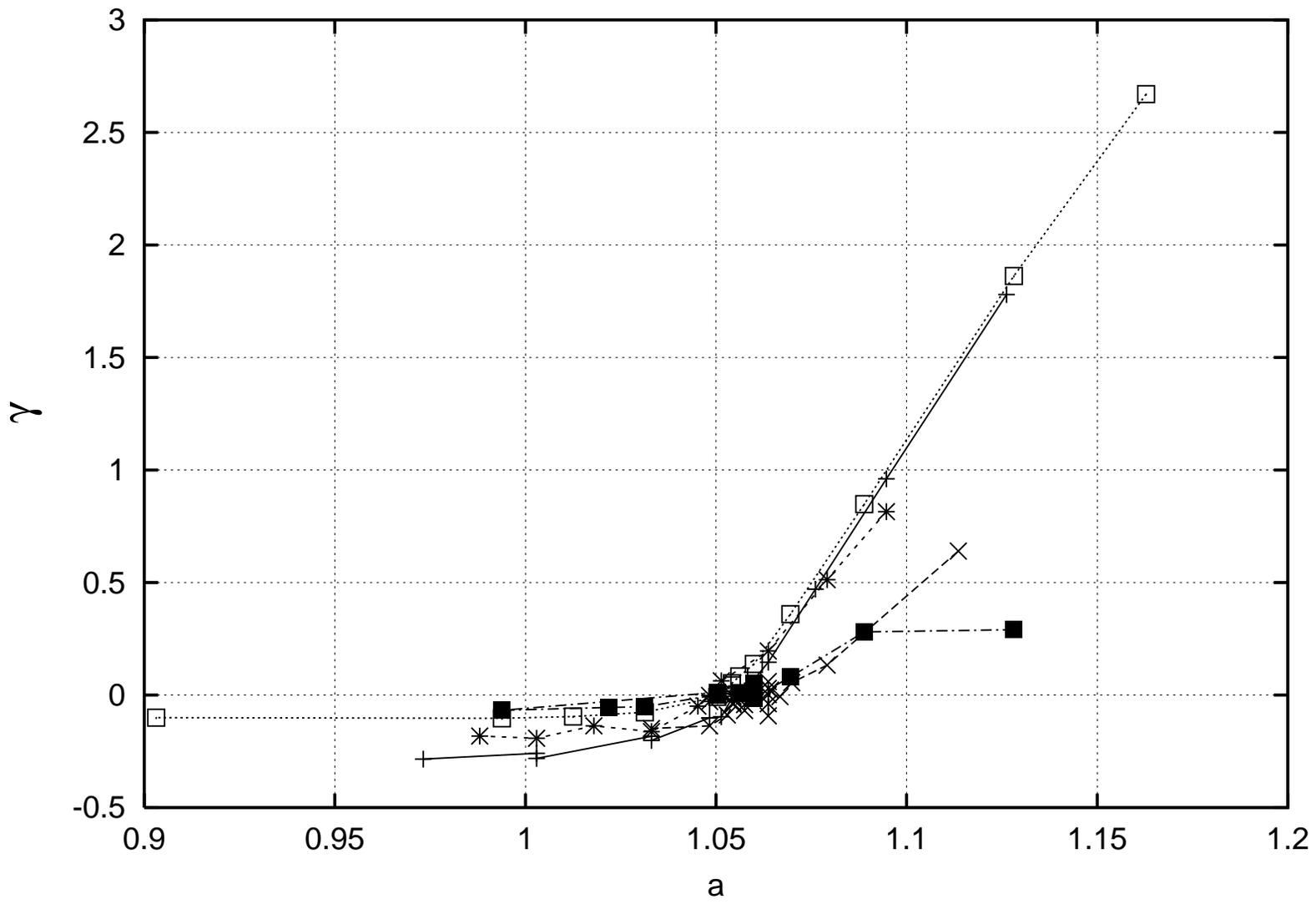
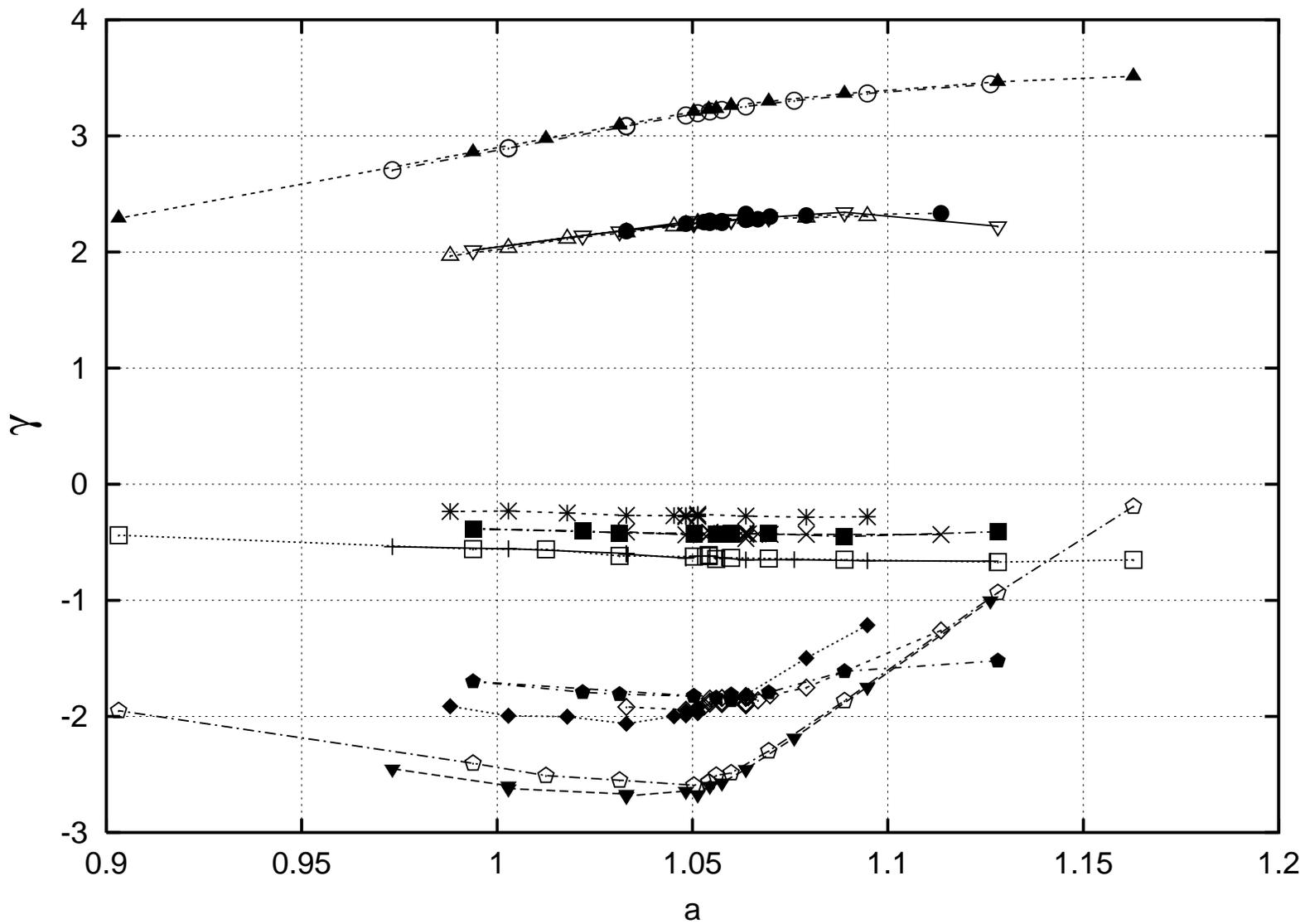
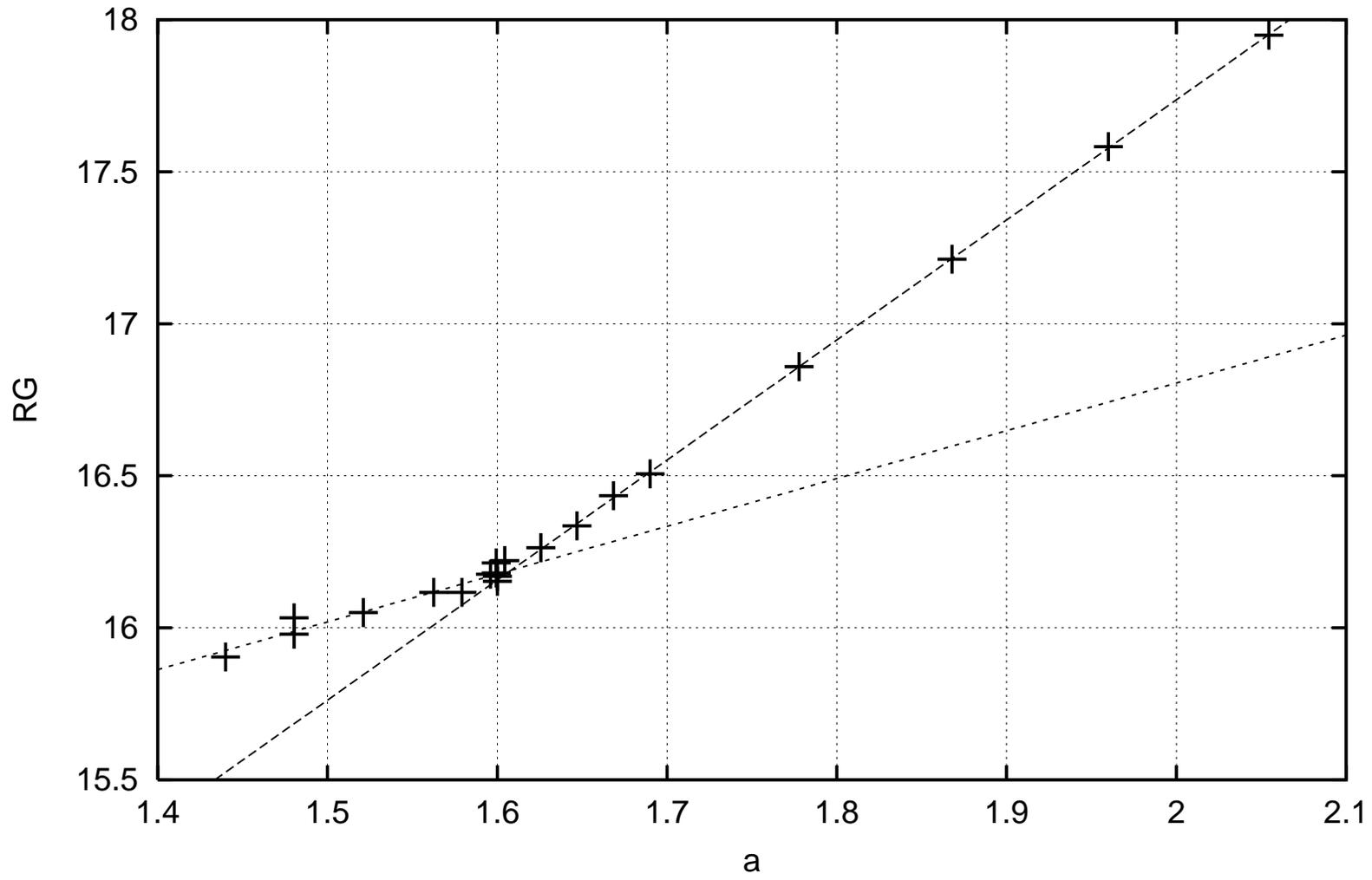
\bye